\documentclass[12pt]{amsart}
\usepackage{amsmath}%,amsthm}%,amssymb
\usepackage{amscd}
\usepackage[psamsfonts]{amssymb}
\usepackage[mathscr,psamsfonts]{eucal}

\IfFileExists{Ueur.fd}{\input{Ueur.fd}}{\input{ueur.fd}}
\IfFileExists{Ueur57.fd}{\input{Ueur57.fd}}{\input{ueur57.fd}}
\DeclareMathAlphabet{\eurm}{U}{eur}{m}{n}
\DeclareMathAlphabet{\eubf}{U}{eur}{b}{n}
\DeclareMathAlphabet{\cyrm}{U}{UWCyr}{m}{n}
\DeclareMathAlphabet{\cyit}{U}{UWCyr}{m}{it}
\DeclareMathAlphabet{\cysc}{U}{UWCyr}{m}{sc}
\DeclareMathAlphabet{\cybf}{U}{UWCyr}{b}{n}
\newcommand{\mysec}[1]{\section{#1}}

\newcommand{\myssec}[1]{\subsection{#1}}

\newtheoremstyle
{MyThm}%        Name
{10pt}%         Space above
{10pt}%         Space below
{\itshape}%     Body font
{\parindent}%   Indent amount (empty = no indent)
{\bfseries}%    Thm head font
{.}%            Punctuation after thm head
{.5em}%         Space after thm head: " " = normal interword space;
{}%             Thm head spec (can be left empty, meaning `normal')
\swapnumbers
\theoremstyle{MyThm}
\newcounter{assump}
\newtheorem{Assumption}{Assumption}[assump]

\newcounter{postul}
\newtheorem{Postulate}{Postulate}[postul]

\newtheorem{Caution}{Caution}[section]
\newtheorem{Convention}[Caution]{Convention}
\newtheorem{Corollary}[Caution]{Corollary}
\newtheorem{Definition}[Caution]{Definition}
\newtheorem{Example}[Caution]{Example}
\newtheorem{Exercise}[Caution]{Exercise}
\newtheorem{Lemma}[Caution]{Lemma}
\newtheorem{Notation}[Caution]{Notation}
\newtheorem{Note}[Caution]{Note}
\newtheorem{Problem}[Caution]{Problem}
\newtheorem{Proposition}[Caution]{Proposition}
\newtheorem{Remark}[Caution]{Remark}
\newtheorem{Theorem}[Caution]{Theorem}
\newcommand{\bAs}{\begin{Assumption}\em}
\newcommand{\eAs}{\end{Assumption}}
\newcommand{\bCa}{\begin{Caution}\em}
\newcommand{\eCa}{\end{Caution}}
\newcommand{\bCr}{\begin{Corollary}\em}
\newcommand{\eCr}{\end{Corollary}}
\newcommand{\bCv}{\begin{Convention}\em}
\newcommand{\eCv}{\end{Convention}}
\newcommand{\bDf}{\begin{Definition}\em}
\newcommand{\eDf}{\end{Definition}}
\newcommand{\bDr}{\begin{Exercise}\em}
\newcommand{\eDr}{\end{Exercise}}
\newcommand{\bEx}{\begin{Example}\em}
\newcommand{\eEx}{\end{Example}}
\newcommand{\bLm}{\begin{Lemma}\em}
\newcommand{\eLm}{\end{Lemma}}
\newcommand{\bNo}{\begin{Notation}\em}
\newcommand{\eNo}{\end{Notation}}
\newcommand{\bNt}{\begin{Note}\em}
\newcommand{\eNt}{\end{Note}}

\newcommand{\bPb}{\begin{Problem}\em}
\newcommand{\ePb}{\end{Problem}}
\newcommand{\bPf}{\begin{proof}[\noindent\indent{\sc Proof}]}
\newcommand{\ePf}{\renewcommand{\qedsymbol}{}\end{proof}}
\newcommand{\bpf}{\bfz\bPf}
\newcommand{\epf}{\ePf\efz}
\newcommand{\bPr}{\begin{Proposition}\em}
\newcommand{\ePr}{\end{Proposition}}
\newcommand{\bPs}{\begin{Postulate}\em}
\newcommand{\ePs}{\end{Postulate}}
\newcommand{\bRm}{\begin{Remark}\em}
\newcommand{\eRm}{\end{Remark}}

\newcommand{\bTh}{\begin{Theorem}}
\newcommand{\eTh}{\end{Theorem}}
\newcommand{\bEq}{\begin{eqnarray}}
\newcommand{\eEq}{\end{eqnarray}}
\newcommand{\beq}{\begin{eqnarray*}}
\newcommand{\eeq}{\end{eqnarray*}}
\newcommand{\bal}{\begin{align*}}
\newcommand{\bAl}{\begin{align}}
\newcommand{\bat}{\begin{alignat*}}
\newcommand{\bAt}{\begin{alignat}}
\newcommand{\bml}{\begin{multline*}}
\newcommand{\bMl}{\begin{multline}}
\newcommand{\bgt}{\begin{gather*}}
\newcommand{\bGt}{\begin{gather}}
\newcommand{\bCd}{\bEq\begin{CD}}
\newcommand{\eCd}{\end{CD}\eEq}
\newcommand{\bcd}{\beq\begin{CD}}
\newcommand{\ecd}{\end{CD}\eeq}
\newcommand{\bdg}{\beq\begin{diagram}}
\newcommand{\edg}{\end{diagram}\eeq}
\newcommand{\bDg}{\bEq\begin{diagram}}
\newcommand{\eDg}{\end{diagram}\eEq}
\newcommand{\bmt}{\left(\begin{matrix}}
\newcommand{\emt}{\end{matrix}\right)}
\newcommand{\bcn}{\begin{center}}
\newcommand{\ecn}{\end{center}}
\newcommand{\ben}{\begin{enumerate}}
\newcommand{\een}{\end{enumerate}}
\newcommand{\btb}{\begin{tabbing}}
\newcommand{\etb}{\end{tabbing}}
\newcommand{\bsm}{\begin{quotation}\small}
\newcommand{\esm}{\end{quotation}}
\newcommand{\bfz}{\begin{footnotesize}}
\newcommand{\efz}{\end{footnotesize}}
\newcommand{\bsz}{\begin{scriptsize}}
\newcommand{\esz}{\end{scriptsize}}
\newcommand{\bsb}
{\vspace{-0.8cm}
\begin{alignat*}{2}
& \qquad\qquad\qquad\qquad\qquad\qquad\qquad\qquad\qquad\qquad
&&\qquad\qquad\qquad\qquad\qquad\qquad\qquad\qquad\qquad
\\}
\newcommand{\Rn}{{I\!\!R}}
\newcommand{\Cn}{{\B C}}

\newcommand{\Al}{\forall}

\newcommand{\coi}{{\mathfrak{i}\,}}

\newcommand{\der}{\partial}

\newcommand{\nab}{\nabla}
\newcommand{\mto}{\mapsto}

\newcommand{\sub}{\subset}

\newcommand{\com}{\circ}

\newcommand{\car}{\times}
\newcommand{\ten}{\otimes}

\newcommand{\wed}{\wedge}

\DeclareMathOperator{\con}{\lrcorner}

\newcommand{\eqv}{\,\equiv\,}
\newcommand{\seq}{\,\simeq\,}

\DeclareMathOperator{\byd}{\,{\rm =}{\raisebox{.092ex}{\rm :}}\,}
\newcommand{\ucar}[1]{\underset{#1}{\times}}

\newcommand{\tfr}[2]{\tfrac{#1}{#2}\,}
\newcommand{\ENDE}{{\,\text{\footnotesize\qedsymbol}}}
\newcommand{\QED}{{\,\text{\rm{\footnotesize QED}}}}

\newcommand{\ssep}[1]{{\qquad\text{\rm{#1}}\qquad}}

\newcommand{\ar}[1]{\url{http://arXiv.org/abs/#1}}

\newcommand{\her}{{{}{\rm her \, }}}

\newcommand{\lin}{{{}{\rm lin \, }}}

\DeclareMathOperator{\id}{{{id}}}
\DeclareMathOperator{\map}{{{map}}}

\DeclareMathOperator{\proj}{{{proj}}}

\DeclareMathOperator{\tr}{{{tr}}}

\newcommand{\f}[1]{{\boldsymbol{#1}}}

\newcommand{\ul}[1]{{\underline{#1}}}

\newcommand{\ba}[1]{{{\bar{#1}}}}

\newcommand{\ch}[1]{{\check{#1}}}
\newcommand{\wch}[1]{{\overset{\vee}{#1}}}

\newcommand{\ti}[1]{{\tilde{#1}}}

\newcommand{\br}[1]{\breve{#1}{}}

\newcommand{\bma}{\left(\begin{matrix}}
\newcommand{\ema}{\end{matrix}\right)}
\newcommand{\R}[1]{{{\rm{#1}}}}

\newcommand{\E}[1]{{\eurm{#1}}}

\newcommand{\F}[1]{{\mathfrak{#1}}}

\newcommand{\B}[1]{{\mathbb{#1}}}

\newcommand{\alp}{\alpha}
\newcommand{\bet}{\beta}

\newcommand{\lam}{\lambda}

\newcommand{\sig}{\sigma}

\newcommand{\Lam}{\Lambda}

\newcommand{\Sig}{\Sigma}

\begin{document}
%--------------------------------------------------------------------%

\title[Graded Lie algebra of Hermitian tangent valued
forms]{Graded Lie algebra of\\Hermitian tangent valued
forms}
\author[J. Jany\v{s}ka, M. Modugno]{
Josef Jany\v{s}ka$^1$, Marco Modugno$^2$}

\address{
{}
\newline
$^1$ Department of Mathematics, Masaryk University
\newline
Jan\'a\v{c}kovo n\'am 2a, 662 95 Brno, Czech Republic
\newline
email: {\tt janyska@math.muni.cz}
\newline
{ }
\newline
$^2$ Department of Applied Mathematics, Florence University
\newline
Via S. Marta 3, 50139 Florence, Italy
\newline
email: {\tt marco.modugno@unifi.it}
}



\thanks{This research has been supported
by Ministry of Education of the Czech Republic under the project
MSM0021622409,
by the Grant agency of the Czech Republic under the project GA
201/05/0523,
by MIUR of Italy under the project PRIN 2003 ``Sistemi integrabili,
teorie classiche e quantistiche'',
by GNFM of INdAM
and by Florence University.}

%\date{}
%\edition{\sm \emph{Preprint: 2005.04.14. - 17.20.}}
%\pagestyle{headings}
%--------------------------------------------------------------------%
\maketitle
%--------------------------------------------------------------------%
\begin{abstract}
We define the Hermitian tangent valued forms of a complex
1--dimensional line bundle equipped with a Hermitian metric.
We provide a local characterisation of these forms in terms of a
local basis and of a local fibred chart.
We show that these forms constitute a graded Lie algebra through the
Fr\"olicher--Nijenhuis bracket.

Moreover, we provide a global characterisation of this graded Lie
algebra, via a given Hermitian connection, in terms of a graded
Lie algebra which is generated by tangent valued forms and forms of
the base space and which involved the curvature of the given
Hermitian connection.
\end{abstract}

\smallskip

%--------------------------------------------------------------------%
%\newpage
\tableofcontents
%--------------------------------------------------------------------%
\section*{Introduction}
\label{Introduction}
%--------------------------------------------------------------------%
In the theory of so called ``Covariant Quantum Mechanics"
(see, for instance, \cite{CanJadMod95,JadMod92,JanMod97b})
a basic role is played by Hermitian vector fields on a complex line
bundle in the frameworks of Galilei and Einstein spacetimes.
In fact, it has been proved that the Lie algebra of Hermitian vector
fields is naturally isomorphic to a Lie algebra of ``special
functions" of the phase space.
Indeed, this is the source of the covariant quantisation of the above
special functions. In the original version of the theory, this result
was formulated and proved in a rather involved way; now, we have
achieved a more direct and simple approach to the classification of
Hermitian vector fields and to their representation via special phase
functions.

In view of a possible covariant quantisation of a larger class of
``observables"
\cite{JanMod99},
it is natural to consider the Hermitian tangent valued forms.

Thus, this paper is devoted to a self--contained analysis of the
graded Lie algebra of Hermitian tangent valued forms of a complex
line bundle and to their classification in terms of tangent valued
forms and forms of the base space.
The local classification is obtained in coordinates. For the global
classification we need a Hermitian connection: indeed, this is just
the connection required in gauge theories.

\smallskip

All manifolds and maps between manifolds are supposed to be smooth.

If
$\f M$
and
$\f N$
are manifolds, and
$\f F \to \f B$
is a fibred manifold, then the sheaf of local smooth maps
$\f M \to \f N$
is denoted by
$\map(\f M, \, \f N) \,,$
the sheaf of local sections
$\f B \to \f F$
is denoted by
$\sec(\f B, \, \f F)$
and the vertical restriction of forms will be denoted by the check
symbol
$^\wch{\,} \,.$
%--------------------------------------------------------------------%
%\newpage
\mysec{Hermitian line bundle}
\label{Hermitian line bundle}
%--------------------------------------------------------------------%
\bsm
We start with some basic properties of a Hermitian line bundle.
\esm

Let us consider a manifold
$\f E \,.$
The charts of
$\f E$
are denoted by
$(x^\lam)$
and the associated local bases of vector fields and forms by
$\der_\lam$
and
$d^\lam \,,$
respectively.

Then, we consider a {\em Hermitian line bundle\/}
$\pi : \f Q \to \f E \,,$
i.e. a complex vector bundle with 1-dimensional fibres, equipped with
a Hermitian product
\cite{GreHalVan72}
$\E h : \f E \to \Cn \ten (\f Q^* \ten \f Q^*) \,.$

The tensor product symbol
$\ten$
always indicates a real tensor product.

We shall refer to {\em quantum bases\/}, i.e. to (local) sections
$\E b \in \sec(\f E,\f Q) \,,$
such that
$\E h(\E b, \E b) = 1$
and to the associated complex dual functions
$z \in \map(\f Q, \Cn) \,.$

We shall also refer to the associated (local) real basis
$(\E b_\R a) \eqv (\E b_1, \E b_2) \byd (\E b, \coi \E b)$
and to the associated scaled real dual basis
$(w^\R a) \eqv (w^1, \, w^2)
= \big(\tfr12 (z + \ba z), \, \tfr12 \coi (\ba z - z)\big) \,.$
We denote the associated vertical vector fields by
$(\der_\R a) \eqv (\der_1, \der_2) \,.$

The small Latin indices
$\R a, \R b = 1, 2$
will span the real indices of the fibres.

Thus, for each
$\Psi \in \sec(\f E, \f Q) \,,$
we write
\bml
\Psi = \Psi^\R a \, \E b_\R a = \psi \, \E b \,,
\ssep{with}
\Psi^1, \Psi^2 \in \map(\f E, \Rn) \,,
\quad
\psi = \Psi^1 + \coi \Psi^2  \in \map(\f E, \Cn)
\end{multline*}
and, for each
$\Phi, \Psi \in \sec(\f E, \f Q) \,,$
\beq
\E h(\Phi, \Psi) =
(\Phi^1 \, \Psi^1 + \Phi^2 \, \Psi^2) + \coi
(\Phi^1 \, \Psi^2 - \Phi^2 \, \Psi^1) =
\ba\phi \, \psi \,.
\eeq

Each
$\Psi \in \sec(\f E, \, \f Q)$
can be naturally regarded as the vertical vector field
\beq
\Psi \seq \ti\Psi \in \sec(\f Q, V\f Q) :
q_e \mto \big(q_e, \Psi(e) \big) \,,
\eeq
with coordinate expression
\beq
\Psi \seq \ti\Psi = \Psi^\R a \, \der_\R a \,.
\eeq

We can regard
$\E h$
also as a complex vertical valued form
$\E h : \f Q \to \Cn \ten V^*\f Q \,,$
with coordinate expression
$\E h =
(w^1 \, \ch d^1 + w^2 \, \ch d^2) + \coi
(w^1 \, \ch d^2 - w^2 \, \ch d^1) \,.$

The {\em unity\/} and the {\em imaginary unity\/} tensors
\beq
1 = \id_\f Q : \f E \to \f Q^* \ten \f Q
\ssep{and}
\coi = \coi \id_\f Q : \f E \to \f Q^* \ten \f Q
\eeq
will be naturally identified, respectively, with the {\em Liouville\/}
and the {\em imaginary Liouville\/} vector fields
\bml
\B I : \f Q \to V\f Q = \f Q \ucar{\f E} \f Q : q \mto (q, q)
\ssep{and}
\coi\B I : \f Q \to V\f Q = \f Q \ucar{\f E} \f Q : q \mto (q, \coi q)
\,.
\end{multline*}

We have the coordinate expressions
\bat{6}
1
&= \id_\f Q
&&= w^\R 1 \, \E b_\R 1 + w^\R 2 \, \E b_\R 2
&&= z \ten \E b \,,
\qquad
&&\B I
&&= w^1 \, \der_1 + w^2 \, \der_2
&&= z \ten \der_1
\\
\coi
&= \coi \id_\f Q
&&= w^1 \, \E b_2 - w^2 \, \E b_1
&&= \coi z \ten \E b \,,
\qquad
&&\coi \B I
&&= w^1 \, \der_2 - w^2 \, \der_1
&&= \coi z \ten \der_1 \,.
\end{alignat*}
%--------------------------------------------------------------------%
\mysec{Tangent valued forms}
\label{Tangent valued forms}
%--------------------------------------------------------------------%
\myssec{Tangent valued forms of a manifold}
\label{Tangent valued forms of a manifold}
%--------------------------------------------------------------------%
\bsm
First of all, we summarise a few essential recalls on tangent valued
forms of a manifold.
\esm

Let us consider a manifold
$\f M$
and denote a generic chart by
$(x^\lam) \,.$

For each integer
$0 \leq r \,,$
let us consider the sheaf
$\sec(\f M, \, \Lam^r T^*\f M \ten T\f M)$
of {\em tangent valued forms\/} of degree
$r \,.$
In particular, for
$r = 0 \,,$
we have the sheaf
$\sec(\f M, T\f M)$
of vector fields.

Let us consider a
$\Xi \in \sec(\f M, \, \Lam^r T^*\f M \ten T\f M) \,.$
Then, we obtain the derivations
\cite{KolMicSlo93}
\bal
i(\Xi)
&:
\sec(\f M, \, \Lam^s T^*\f M) \to
\sec(\f M, \, \Lam^{r+s-1} T^*\f M)
\\
L(\Xi)
&:
\sec(\f M, \, \Lam^s T^*\f M) \to
\sec(\f M, \, \Lam^{r+s} T^*\f M) \,,
\end{align*}
which are characterised, via decomposable tangent valued forms, by the
equalities
\bal
i(\xi \ten X) \, \alp
&=
\xi \wed i(X) \, \alp
\\
L(\xi \ten X) \, \alp
&=
\xi \wed L(X) \, \alp - (-1)^{r-1} d\xi \wed i(X) \, \alp \,.
\end{align*}

\smallskip

We have the natural (real) linear injective morphisms
\bal
\sec(\f E, \, \Lam^r T^*\f E) \to
\sec(\f Q, \, \Lam^r T^*\f E \ten V\f Q)
&:
\xi \mto \xi \ten \B I
\\
\sec(\f E, \, \Lam^r T^*\f E) \to
\sec(\f Q, \, \Lam^r T^*\f E \ten V\f Q)
&:
\xi \mto \coi \xi \ten \B I \,,
\end{align*}
whose inverse are, respectively,
\beq
\tr_\Cn : \xi \ten \B I \mto \xi
\ssep{and}
- \coi \tr_\Cn : \xi \ten \B I \mto \xi \,.
\eeq

The sheaf of tangent valued forms turns out to be a graded Lie
algebra with respect to the Fr\"olicher-Nijenhuis bracket (FN bracket)
\cite{KolMicSlo93}
\beq
\sec(\f M, \, \Lam^r T^*\f M \ten T\f M) \car
\sec(\f M, \, \Lam^s T^*\f M \ten T\f M) \to
\sec(\f M, \, \Lam^{r+s} T^*\f M \ten T\f M) \,,
\eeq
which is characterised, via decomposable tangent valued forms, by the
equality
\bml
[\xi \ten X, \; \sig \ten Y] =
\xi \wed \sig \ten [X, Y] +
\xi \wed L(X) \, \sig \ten Y - (-1)^{rs}
\sig \wed L(Y) \, \xi \ten X
\\
+ (-1)^r
d\xi \wed i(X) \, \sig \ten Y - (-1)^{s + rs}
d\sig \wed i(Y) \, \xi \ten X \,.
\end{multline*}

We have the coordinate expression
\bml
[\Xi, \, \Sig]
= (\Xi^\rho_{\lam_1 \dots \lam_r} \,
\der_\rho \Sig^\mu_{\lam_{r+1} \dots \lam_{r+s}}
- (- 1)^{rs} \,
\Sig^\rho_{\lam_1 \dots \lam_s} \,
\der_\rho \Xi^\mu_{\lam_{s+1} \dots \lam_{r+s}}
\\
- r \, \Xi^\mu_{\lam_1 \dots \lam_{r-1} \rho} \,
\der_{\lam_r} \Sig^\rho_{\lam_{r+1} \dots \lam_{r+s}}
+ (- 1)^{rs} \, s \,
\Sig^\mu_{\lam_1 \dots \lam_{s-1} \rho} \,
\der_{\lam_s} \Xi^\rho_{\lam_{s+1} \dots \lam_{r+s}}) \,
d^{\lam_1} \wed \dots \wed d^{\lam_{r+s}} \ten \der_\mu
\,.
\end{multline*}

We have the identity
\beq
[L(\Xi), \, L(\Sig)] \byd
L(\Xi) \com L(\Sig) - (-1)^{rs} L(\Sig) \com L(\Xi) =
L\big([\Xi, \Sig]\big) \,.
\eeq
%--------------------------------------------------------------------%
\myssec{Projectable tangent valued forms}
\label{Projectable tangent valued forms}
%--------------------------------------------------------------------%
\bsm
Now, we analyse a distinguished subsheaf of the tangent valued forms
of the line bundle.
\esm

Let us devote our attention to the sheaf
$\sec(\f Q, \, \Lam^r T^*\f E \ten T\f Q) \,.$

The coordinate expression of
$\Xi \in \sec(\f Q, \, \Lam^r T^*\f E \ten T\f Q)$
is of the type
\bal
\Xi
&=
d^{\lam_1} \wed \dots \wed d^{\lam_r} \ten
(\Xi^\mu_{\lam_1 \dots \lam_r} \, \der_\mu +
\Xi^\R a_{\lam_1 \dots \lam_r} \, \der_\R a)
\\
&=
d^{\lam_1} \wed \dots \wed d^{\lam_r} \ten
(\Xi^\mu_{\lam_1 \dots \lam_r} \, \der_\mu +
\Xi^z_{\lam_1 \dots \lam_r} \, \der_1) \,,
\end{align*}
with
$\Xi^\mu_{\lam_1 \dots \lam_r}, \Xi^\R a_{\lam_1 \dots \lam_r} \in
\map(\f Q, \Rn)$
and
$\Xi^z_{\lam_1 \dots \lam_r} =
\Xi^1_{\lam_1 \dots \lam_r} + \coi \Xi^2_{\lam_1 \dots \lam_r} \,.$

\smallskip

$\Xi$
is said to be {\em projectable\/}
if \,
$T\pi \com \Xi \in \sec(\f Q, \, \Lam^r T^*\f E \ten T\f E)$
factorises through a section
$\ul \Xi \in \sec(\f E, \, \Lam^r T^*\f E \ten T\f E) \,.$

Thus,
$\Xi$
is projectable if and only if
$\Xi^\mu_{\lam_1 \dots \lam_r} \in \map(\f E, \Rn) \,.$

We denote the subsheaf of projectable tangent valued forms of
degree
$r$
by
\beq
\proj(\f Q, \, \Lam^r T^*\f E \ten T\f Q) \sub
\sec(\f Q, \, \Lam^r T^*\f E \ten T\f Q) \,.
\eeq

In particular, we have the subsheaf of {\em vertical valued forms\/}
\beq
\sec(\f Q, \, \Lam^r T^*\f E \ten V\f Q) \sub
\proj(\f Q, \, \Lam^r T^*\f E \ten T\f Q) \,.
\eeq

\smallskip

The sheaf of projectable tangent valued forms is closed with respect
to the FN bracket.

For projectable tangent valued forms, we have the identity
\beq
[\ul\Xi, \ul\Sig] = \ul{[\Xi, \Sig]} \,.
\eeq

For projectable tangent valued forms, we obtain the coordinate
expression
\bal
[\Xi, \Sig]
&= (\Xi^\rho_{\lam_1 \dots \lam_r} \,
\der_\rho \Sig^\mu_{\lam_{r+1} \dots \lam_{r+s}}
- (- 1)^{rs} \,
\Sig^\rho_{\lam_1 \dots \lam_s} \,
\der_\rho \Xi^\mu_{\lam_{s+1} \dots \lam_{r+s}}
\\
&- r \, \Xi^\mu_{\lam_1 \dots \lam_{r-1} \rho} \,
\der_{\lam_r} \Sig^\rho_{\lam_{r+1} \dots \lam_{r+s}}
+ (- 1)^{rs} \, s \,
\Sig^\mu_{\lam_1 \dots \lam_{s-1} \rho} \,
\der_{\lam_s} \Xi^\rho_{\lam_{s+1} \dots \lam_{r+s}}) \cdot
\\
&\quad\quad\quad\qquad\quad\quad\quad\qquad\quad
\quad\quad\qquad\quad\quad\quad\qquad\qquad\qquad
\cdot d^{\lam_1} \wed \dots \wed d^{\lam_{r+s}} \ten \der_\mu
\\
&+ (\Xi^\rho_{\lam_1 \dots \lam_r} \,
\der_\rho \Sig^\R a_{\lam_{r+1} \dots \lam_{r+s}}
- (- 1)^{rs} \,
\Sig^\rho_{\lam_1 \dots \lam_s} \,
\der_\rho \Xi^\R a_{\lam_{s+1} \dots \lam_{r+s}}
\\
&+ \Xi^\R b_{\lam_1 \dots \lam_r} \,
\der_\R b \Sig^\R a_{\lam_{r+1} \dots \lam_{r+s}}
- (- 1)^{rs} \,
\Sig^\R b_{\lam_1 \dots \lam_s} \,
\der_\R b \Xi^\R a_{\lam_{s+1} \dots \lam_{r+s}}
\\
&- r \, \Xi^\R a_{\lam_1 \dots \lam_{r-1} \rho} \,
\der_{\lam_r} \Sig^\rho_{\lam_{r+1} \dots \lam_{r+s}}
+ (- 1)^{rs} \, s \,
\Sig^\R a_{\lam_1 \dots \lam_{s-1} \rho} \,
\der_{\lam_s} \Xi^\rho_{\lam_{s+1} \dots \lam_{r+s}}) \cdot
\\
&\quad\quad\quad\qquad\quad\quad\quad\qquad\quad
\quad\quad\qquad\quad\quad\quad\qquad\qquad\qquad
\cdot d^{\lam_1} \wed \dots \wed d^{\lam_{r+s}} \ten \der_\R a
\,.
\end{align*}

Moreover, for decomposable projectable tangent valued forms, we obtain
\bml
[\xi \ten X, \; \sig \ten Y] =
\xi \wed \sig \ten [X, Y] + \xi \wed L_\ul X \sig \ten Y
- (-1)^{rs} \, \sig \wed L_\ul Y \xi \ten X
\\
+ (- 1)^{r} \, d\xi \wed i_\ul X \sig \ten Y
- (- 1)^{rs + s} \, d\sig \wed i_\ul Y\xi \ten X \,.
\end{multline*}
%--------------------------------------------------------------------%
\myssec{Linear tangent valued forms}
\label{Linear tangent valued forms}
%--------------------------------------------------------------------%
\bsm
Next, we analyse the subsheaf of linear tangent valued forms of the
line bundle.
\esm

A projectable tangent valued form
$\Xi$
is said to be (real) {\em linear\/} if it is a (real) linear fibred
morphism over its projection
$\ul\Xi \,.$

Thus, a projectable tangent valued form
$\Xi$
is (real) linear if and only if
\beq
\Xi^\R a_{\lam_1 \dots \lam_r} =
\Xi^\R a_{\lam_1 \dots \lam_r \, \R b} \, w^\R b \,,
\ssep{with}
\Xi^\R a_{\lam_1 \dots \lam_r \, \R b} \in \map(\f E, \Rn) \,.
\eeq

If
$\Xi$
is a (real) linear tangent valued form, then we have
$[\Xi, \B I] = 0 \,.$

We denote the subsheaf of (real) linear tangent valued forms of
degree
$r$
by
\beq
\lin_\Rn(\f Q, \, \Lam^r T^*\f E \ten T\f Q) \sub
\proj(\f Q, \, \Lam^r T^*\f E \ten T\f Q) \,.
\eeq

The sheaf of (real) linear tangent valued forms is closed with respect
to the FN bracket.

\smallskip

A (real) linear tangent valued form
$\Xi$
is said to be complex {\em linear\/} if it is a complex linear fibred
morphism over its projection
$\ul\Xi \,.$

Thus, a projectable tangent valued form
$\Xi$
is complex linear if and only if
\beq
\Xi^z_{\lam_1 \dots \lam_r} =
\Xi^z_{\lam_1 \dots \lam_r \, z} \, z \,,
\ssep{with}
\Xi^z_{\lam_1 \dots \lam_r \, z} \in \map(\f E, \Rn) \,,
\eeq
i.e., if and only if
\beq
\Xi^1_{\lam_1 \dots \lam_r \, 1} = \Xi^2_{\lam_1 \dots \lam_r \, 2}
\ssep{and}
\Xi^2_{\lam_1 \dots \lam_r \, 1} = -\Xi^1_{\lam_1 \dots \lam_r \, 2}
\,.
\eeq

In such a case, we have
\beq
\Xi^z_{\lam_1 \dots \lam_r \, z} =
\Xi^1_{\lam_1 \dots \lam_r \, 1} + \coi
\Xi^2_{\lam_1 \dots \lam_r \, 1} =
\Xi^2_{\lam_1 \dots \lam_r \, 2} - \coi
\Xi^1_{\lam_1 \dots \lam_r \, 2} \,.
\eeq

If
$\Xi$
is a complex linear tangent valued form, then we have
$[\Xi, \coi \B I] = 0 \,.$

We denote the subsheaf of complex linear tangent valued forms of
degree
$r$
by
\beq
\lin_\Cn(\f Q, \, \Lam^r T^*\f E \ten T\f Q) \sub
\lin_\Rn(\f Q, \, \Lam^r T^*\f E \ten T\f Q) \,.
\eeq

The sheaf of complex linear tangent valued forms is closed with
respect to the FN bracket.
%--------------------------------------------------------------------%
\myssec{Hermitian tangent valued forms}
\label{Hermitian tangent valued forms}
%--------------------------------------------------------------------%
\bsm
Eventually, we introduce the notion of Hermitian tangent valued forms.
\esm

\bLm
If
$\alp \in \sec(\f Q, V^*\f Q)$
and
$\Xi \in \proj(\f Q, \, \Lam^rT^*\f E \ten T\f Q) \,,$
then the Lie derivative
$L(\Xi) \alp$
is well defined, in spite of the fact that the form
$\alp$
is vertical valued, and has coordinate expression
\beq
L(\Xi) \alp = (
\Xi^\mu_{\lam_1 \dots \lam_r} \, \der_\mu \alp_\R a +
\Xi^\R b_{\lam_1 \dots \lam_r} \, \der_\R b \, \alp_\R a +
\alp_\R b \, \der_\R a \, \Xi^\R b_{\lam_1 \dots \lam_r}) \,
d^{\lam_1} \wed \dots \wed d^{\lam_r} \ten \ch d^\R a \,.
\eeq
\eLm

\bpf
If
$\ti\alp \in \sec(\f Q, T^*\f Q)$
is any extension of
$\alp$
(obtained, for instance through a connection of the line bundle), then
let us prove that the vertical restriction ``to one variable"
\beq
L(\Xi) \alp \byd (L(\Xi) \ti\alp)^{\wch{\,}_1} \in
\sec(\f Q, \, \Lam^rT^*\f E \ten V^*\f Q)
\eeq
does not depend on the choice of the extension
$\ti\alp \,.$

The coordinate expression of
$\ti\alp$
is of the type
$\ti\alp = \alp_\mu \, d^\mu + \alp_\R a \, d^\R a \,.$

Then, the expression
$\Xi = d^{\lam_1} \wed \dots \wed d^{\lam_r} \ten
(\Xi^\lam_{\lam_1 \dots \lam_r} \, \der_\lam +
\Xi^\R a_{\lam_1 \dots \lam_r} \, \der_\R a) \,,$
with
$\der_\R b \, \Xi^\lam_{\lam_1 \dots \lam_r} = 0 \,,$
yields
\bml
L(\Xi) \, \ti\alp =
d^{\lam_1} \wed \dots \wed d^{\lam_r} \wed
\\
\wed \big((
\Xi^\mu_{\lam_1 \dots \lam_r} \, \der_\mu \alp_\lam +
\Xi^\R b_{\lam_1 \dots \lam_r} \, \der_\R b \, \alp_\lam +
\alp_\mu \, \der_\lam \Xi^\mu_{\lam_1 \dots \lam_r} +
\alp_\R b \, \der_\lam \Xi^\R b_{\lam_1 \dots \lam_r} ) \, d^\lam
\\
+ (
\Xi^\mu_{\lam_1 \dots \lam_r} \, \der_\mu \alp_\R a +
\Xi^\R b_{\lam_1 \dots \lam_r} \, \der_\R b \, \alp_\R a +
\alp_\R b \, \der_\R a \, \Xi^\R b_{\lam_1 \dots \lam_r}) \,
d^\R a\big)
\,.
\end{multline*}

Eventually, by considering the natural map
\beq
^{\wch{\,}_1} : \ten^{r+1}T^*\f Q \to
\ten^rT^*\f Q \ten V^*\f Q :
\bet^1 \ten \dots \ten \bet^{r+1} \mto
\sum_{1 \leq i \leq r+1} \,
\bet^1 \ten \dots \ten \ch\bet{}^i \ten \dots \ten \bet^{r+1}\,,
\eeq
we obtain the section
\beq
\big(L(\Xi) \, \ti\alp\big)^{\wch{\,}_1} = (
\Xi^\mu_{\lam_1 \dots \lam_r} \, \der_\mu \alp_\R a +
\Xi^\R b_{\lam_1 \dots \lam_r} \, \der_\R b \alp_\R a +
\der_\R a \Xi^\R b_{\lam_1 \dots \lam_r} \, \alp_\R b) \,
d^{\lam_1} \wed \dots \wed d^{\lam_r} \ten \ch d^\R a \,,
\eeq
which turns out to be valued in the subspace
$\Lam^rT^*\f E \ten V^*\f Q \sub
\ten^rT^*\f Q \ten V^*\f Q \,.$\QED
\epf

A (real) linear tangent valued form
$\Xi$
is said to be {\em Hermitian\/} if
$L(\Xi) \E h = 0 \,.$

\bLm\label{Lie derivative of h}
For each
$\Xi \in \lin_\Rn(\f Q, \, \Lam^rT^*\f E \ten T\f Q) \,,$
we have the coordinate expression
\begin{gather*}
L(\Xi) \, \E h =
\\
\begin{align*}
&=
\big(2 \, \Xi^1_{\lam_1 \dots \lam_r \, 1} \, w^1 +
(\Xi^2_{\lam_1 \dots \lam_r \, 1} + \Xi^1_{\lam_1 \dots \lam_r \, 2})
\, w^2
- \coi \Xi^\R a_{\lam_1 \dots \lam_r \, \R a} \, w^2
\big) \,
d^{\lam_1} \wed \dots \wed d^{\lam_r} \ten \ch d^1
\\
&+
\big(2 \, \Xi^2_{\lam_1 \dots \lam_r \, 2} \, w^2 +
(\Xi^2_{\lam_1 \dots \lam_r \, 1} + \Xi^1_{\lam_1 \dots \lam_r \, 2})
\, w^1
+ \coi \Xi^\R a_{\lam_1 \dots \lam_r \, \R a} \, w^1
\big) \,
d^{\lam_1} \wed \dots \wed d^{\lam_r} \ten \ch d^2 \,.\ENDE
\end{align*}
\end{gather*}
\eLm

\bPr\label{expression of Hermitian tangent valued forms}
Each Hermitian tangent valued form
$\Xi$
turns out to be complex linear.
Moreover,
$\Xi \in \lin_\Rn(\f Q, \, \Lam^rT^*\f E \ten T\f Q)$
is Hermitian if and only if it is (locally) of the type
\beq
\Xi = \chi[\E b] (\ul\Xi) + \coi \br\Xi[\E b] \ten \B I \,,
\ssep{with}
\br\Xi[\E b] \in \sec(\f E, \, \Lam^rT^*\f E) \,,
\eeq
where
$\chi[\E b]$
is the (local) flat connection of
$\f Q \to \f E$
induced by the basis
$\E b \,.$

In other words,
$\Xi$
is Hermitian if and only if
\beq
\Xi^1_{\lam_1 \dots \lam_r \, 1} =
\Xi^2_{\lam_1 \dots \lam_r \, 2} = 0
\ssep{and}
\Xi^2_{\lam_1 \dots \lam_r \, 1} = -
\Xi^1_{\lam_1 \dots \lam_r \, 2} \,,
\eeq
i.e. if and only if its coordinate expression is of the type
\beq
\Xi = d^{\lam_1} \wed \dots \wed d^{\lam_r} \ten
(\Xi^\lam_{\lam_1 \dots \lam_r} \, \der_\lam +
\coi \, \br \Xi_{\lam_1 \dots \lam_r} \, \B I) \,,
\eeq
with
$\Xi^\lam_{\lam_1 \dots \lam_r} \in \map(\f E, \Rn) \,,
\quad
\br \Xi_{\lam_1 \dots \lam_r} =
\Xi^2_{\lam_1 \dots \lam_r \, 1} = -
\Xi^1_{\lam_1 \dots \lam_r \, 2} \in \map(\f E, \Rn) \,.$\ENDE
\ePr

\bCr
In particular, the Hermitian vertical valued forms
$\Xi$
are of the type
\beq
\Xi = \coi \br\Xi \ten \B I \,,
\ssep{with}
\br\Xi \in \sec(\f E, \, \Lam^rT^*\f E) \,.
\eeq

Hence, the Hermitianity of vertical valued forms does not depend on
the choice of the Hermitian metric
$\E h \,.$
Moreover, the form
$\br\Xi$
is global and does not depend on the choice of the basis
$\E b \,.$\ENDE
\eCr

We denote the subsheaf of Hermitian tangent valued forms of
degree
$r$
by
\beq
\her(\f Q, \, \Lam^r T^*\f E \ten T\f Q) \sub
\lin_\Cn(\f Q, \, \Lam^r T^*\f E \ten T\f Q) \,.
\eeq

Each
$\Xi \in \her(\f Q, \, \Lam^rT^*\f E \ten T\f Q)$
can be written locally as sum of decomposable tangent valued forms
of the type
\beq
\xi \ten Y \,,
\ssep{with}
\xi \in \sec(\f E, \, \Lam^rT^*\f E) \,,
\quad
Y \in \her(\f Q, T\f Q) \,.
\eeq

However, in general this decomposition is not unique and holds only
locally.

If
$\Xi \in \her(\f Q, \, \Lam^r T^*\f E \ten T\f Q)$
and
$\alp \in \sec(\f E, \, \Lam^s T^*\f E) \,,$
then
\beq
\alp \wed \Xi \in
\her(\f Q, \, \Lam^{r+s} T^*\f E \ten T\f Q) \,.
\eeq
%--------------------------------------------------------------------%
\myssec{Graded Lie algebra of Hermitian tangent valued forms}
\label{Graded Lie algebra of Hermitian tangent valued forms}
%--------------------------------------------------------------------%
\bsm
We show that the sheaf of Hermitian tangent valued forms is closed
with respect to the FN bracket.
\esm

\bLm
For each
$\br\Xi \in \sec(\f E, \, \Lam^rT^*\f E)$
and
$\br\Sig \in \sec(\f E, \, \Lam^sT^*\f E)$
we have
\beq
[\coi \br\Xi \ten \B I \,, \;\; \coi \br\Sig \ten \B I] = 0 \,.\ENDE
\eeq
\eLm

\bLm
For each
$\ul\Xi \in \sec(\f E, \, \Lam^rT^*\f E \ten T\f E)$
and
$\ul\Sig \in \sec(\f E, \, \Lam^sT^*\f E \ten T\f E) \,,$
we have
\beq
\big[\chi[\E b] (\ul\Xi) \,, \; \chi[\E b] (\ul\Sig)\big] =
\chi[\E b] \big([\ul\Xi \,, \; \ul\Sig]\big) \,.\ENDE
\eeq
\eLm

\bLm
For each
$\ul\Xi, \in \sec(\f E, \, \Lam^rT^*\f E \ten T\f E)$
and
$\br\Sig \in \sec(\f E, \, \Lam^sT^*\f E) \,,$
we have
\beq
\big[\chi[\E b] (\ul\Xi) \,, \; \coi \br\Sig \ten \B I\big] =
\coi \big(L(\ul\Xi) \, \br\Sig\big) \ten \B I \,.\ENDE
\eeq
\eLm

\bTh
The sheaf of Hermitian tangent valued forms is closed with respect to
the FN bracket.

Indeed, for each
$\Xi \in \her(\f Q, \, \, \Lam^rT^*\f E \ten TQ)$
and
$\Sig \in \her(\f Q, \, \, \Lam^sT^*\f E \ten TQ) \,,$
we have
\bml
\big[\chi[\E b] (\ul\Xi) + \coi \br\Xi[\E b] \ten \B I \,, \;
\chi[\E b] (\ul\Sig) + \coi \br\Sig[\E b] \ten \B I\big] =
\\
=
\chi[\E b] \big([\ul\Xi,\ul\Sig]\big)  +
\coi \big(L(\ul\Xi) \, \br\Sig[\E b] -
(-1)^{rs} L(\ul\Sig) \, \br\Xi[\E b]\big) \ten \B I \,.\ENDE
\end{multline*}
\eTh

\bCr
The sheaf of vertical Hermitian tangent valued forms is an
abelian subalgebra and an ideal of the algebra of Hermitian tangent
valued forms.

Indeed, for each
$\Xi \in \her(\f Q, \, \, \Lam^rT^*\f E \ten TQ)$
and
$\Sig \in \her(\f Q, \, \, \Lam^sT^*\f E \ten VQ) \,,$
we have
\beq
\big[\chi[\E b] (\ul\Xi) + \coi \br\Xi[\E b] \ten \B I \,, \;\;
\coi \br\Sig \ten \B I\big] =
\coi \big(L(\ul\Xi) \, \br\Sig\big) \ten \B I \,.\ENDE
\eeq
\eCr

\bCr
The sheaf of Hermitian vector fields turns out to be a subalgebra of
the algebra of Hermitian tangent valued forms.

Indeed, for each
$X, Y \in \her(\f Q, T\f Q) \,,$
we have
\beq
\big[\chi[\E b] (\ul X) + \coi \br X[\E b] \, \B I \,,\;
\chi[\E b](\ul Y) + \coi \br Y[\E b] \, \B I\big] =
\chi[\E b]([\ul X, \, \ul Y]) +
\coi (\ul X. \br Y[\E b] -\ul Y. \br X[\E b]) \, \B I
\,.\ENDE
\eeq
\eCr

Each
$Y \in \lin_\Rn(\f Q, T\f Q)$
turns out to be Hermitian if and only if
\beq
L(\ul Y) \, \E h(\Phi, \Psi) =
\E h \big(L(Y) \, \Phi \,, \; \Psi\big) +
\E h \big(\Phi \,, \; L(Y) \, \Psi\big) \,,
\qquad
\Al \, \Phi, \Psi \in \sec(\f E, \f Q) \,.
\eeq
%--------------------------------------------------------------------%
\mysec{Classification of Hermitian tangent valued forms}
\label{Classification of Hermitian tangent valued forms}
%--------------------------------------------------------------------%
\bsm
The above results provide a local characterisation of Hermitian
tangent valued forms in terms of a basis or of a fibred chart.

On the other hand, the choice of a global connection allows us to
exhibit a global characterisation of Hermitian tangent valued forms
in terms of tangent valued forms and forms of the base space.
\esm
%--------------------------------------------------------------------%
\myssec{Hermitian connections}
\label{Hermitian connections}
%--------------------------------------------------------------------%
\bsm
In view of the above global characterisation, we recall a few basic
properties of Hermitian connections.
\esm

Let us consider a connection of the line bundle
$c : \f Q \to T^*\f E \ten T\f Q \,,$
i.e., tangent valued 1--form, which is projectable on
$\f1_\f E \,.$

Its coordinate expression is of the type
$c = d^\lam \ten (\der_\lam + c^\R a_\lam \, \der_\R a) \,,$
where
$c^\R a_\lam \in \map(\f Q, \B R) \,.$

The vertical 1--form
$\nu[c] : \f Q \to T^*\f Q \ten V\f Q$
associated with
$c$
has coordinate expression
\beq
\nu[c] =
(d^\R a - c^\R a_\lam \, d^\lam) \ten \der_\R a \,.
\eeq

For each
$\Xi \in \proj(\f E, \, \Lam^rT^*\f E \ten T\f Q) \,,$
we have the {\em covariant exterior differential\/}
\cite{Mod91}
\beq
d[c] \Xi \byd [c, \Xi] :
\proj(\f E, \, \Lam^{r+1}T^*\f E \ten V\f Q)
\,,
\eeq
with coordinate expression
\bml
d[c]\Xi
=
\big(- \der_{\lam_1} \Xi^\rho_{\lam_2 \dots \lam_{r+1}} \,
c^\R a_\rho
- \der_\rho c^\R a_{\lam_1} \,
\Xi^\rho_{\lam_2 \dots \lam_{r+1}}
\\
+ \der_{\lam_1} \Xi^\R a_{\lam_2 \dots \lam_{r+1}}
+ c^\R b_{\lam_1} \,
\der_\R b \Xi^\R a_{\lam_2 \dots \lam_{r+1}}
- \der_\R b c^\R a_{\lam_1} \,
\Xi^\R b_{\lam_2 \dots \lam_{r+1}}\big)
\,  d^{\lam_1} \wed \dots \wed d^{\lam_{r+1}} \ten \der_\R a \,.
\end{multline*}

The {\em curvature\/} of
$c$
is defined to be the vertical valued 2--form
\cite{Mod91}
\beq
R[c] \byd -d[c] c \byd - [c,c] :
\f E \to \Lam^2T^*\f E \ten V\f Q \,,
\eeq
with coordinate expression
$R[c] =
- 2 \, (\der_\lam c^\R a_\mu +
c^\R b_\lam \, \der_\E b c^\R a_\mu) \,
d^\lam \wed d^\mu \ten \der_\R a \,.$

\bLm
For each
$\ul\Xi \in \sec(\f E, \, \Lam^rT^*\f E \ten T\f E)$
and
$\ul\Sig \in \sec(\f E, \, \Lam^sT^*\f E \ten T\f E) \,,$
we have
\beq
[c, \, c(\ul\Xi)] = \ul\Xi \con R[c]
\ssep{and}
[c(\ul\Xi), \, c(\ul\Sig)] =
c\big([\ul\Xi, \, \ul\Sig]\big) - R[c](\ul\Xi, \ul\Sig) \,,
\eeq
where
$\ul\Xi \con R[c]$
and
$R[c](\ul\Xi, \ul\Sig)$
are defined, via decomposable tangent valued forms, by
\bal
(\xi \ten X) \con R[c] &= (-1)^r \xi \wed (X \con R[c])
\\
R[c](\xi \ten X, \, \sig \ten Y) &=
(\xi \wed \sig) \ten (Y \con X \con R[c])
\,.
\end{align*}
\eLm

\bpf
We have
\begin{gather*}
[c, c(\ul\Xi)] =
\\
\begin{align*}
&=
\big(- \der_{\lam_1} \Xi^\rho_{\lam_2 \dots \lam_{r+1}} \,
c^\R a_\rho
- \der_\rho c^\R a_{\lam_1} \,
\Xi^\rho_{\lam_2 \dots \lam_{r+1}}
\\
&+
\der_{\lam_1} \Xi^\rho_{\lam_2 \dots \lam_{r+1}} \, c^\R a_\rho
+ \Xi^\rho_{\lam_2 \dots \lam_{r+1}} \, \der_{\lam_1} c^\R a_\rho
+ c^\R b_{\lam_1} \,
\der_\R b c^\R a_\rho \, \Xi^\rho_{\lam_2 \dots \lam_{r+1}}
- \der_\R b c^\R a_{\lam_1} \,
c^\R b_\rho \, \Xi^\rho_{\lam_2 \dots \lam_{r+1}}\big)
\,  d^{\lam_1} \wed \dots \wed d^{\lam_{r+1}} \ten \der_\R a
\\[3mm]
&=
- (
\der_\rho c^\R a_{\lam_1} -
\der_{\lam_1} c^\R a_\rho +
c^\R b_\rho \, \der_\R b c^\R a_{\lam_1} -
c^\R b_{\lam_1} \, \der_\R b c^\R a_\rho) \,
\Xi^\rho_{\lam_2 \dots \lam_{r+1}}
\,  d^{\lam_1} \wed \dots \wed d^{\lam_{r+1}} \ten \der_\R a
\\
&=
R^\R a_{\rho\lam_1} \,
\Xi^\rho_{\lam_2 \dots \lam_{r+1}}
\,  d^{\lam_1} \wed \dots \wed d^{\lam_{r+1}} \ten \der_\R a \,.
\end{align*}
\end{gather*}

Moreover, we have
\begin{gather*}
[c(\ul\Xi), \, c(\ul\Sig)] =
\\
\begin{align*}
&=
(\Xi^\rho_{\lam_1 \dots \lam_r} \,
\der_\rho \Sig^\mu_{\lam_{r+1} \dots \lam_{r+s}}
- (- 1)^{rs} \,
\Sig^\rho_{\lam_1 \dots \lam_s} \,
\der_\rho \Xi^\mu_{\lam_{s+1} \dots \lam_{r+s}}
\\
&-
r \, \Xi^\mu_{\lam_1 \dots \lam_{r-1} \rho} \,
\der_{\lam_r} \Sig^\rho_{\lam_{r+1} \dots \lam_{r+s}}
+ (- 1)^{rs} \, s \,
\Sig^\mu_{\lam_1 \dots \lam_{s-1} \rho} \,
\der_{\lam_s} \Xi^\rho_{\lam_{s+1} \dots \lam_{r+s}}) \,
d^{\lam_1} \wed \dots \wed d^{\lam_{r+s}} \ten \der_\mu
\\
&+ (\Xi^\rho_{\lam_1 \dots \lam_r} \,
\der_\rho c^\R a_\mu \, \Sig^\mu_{\lam_{r+1} \dots \lam_{r+s}}
+ \Xi^\rho_{\lam_1 \dots \lam_r} \,
c^\R a_\mu \, \der_\rho \Sig^\mu_{\lam_{r+1} \dots \lam_{r+s}}
\\
&- (- 1)^{rs} \, (
\Sig^\rho_{\lam_1 \dots \lam_s} \,
\der_\rho c^\R a_\mu \, \Xi^\mu_{\lam_{s+1} \dots \lam_{r+s}} +
\Sig^\rho_{\lam_1 \dots \lam_s} \,
c^\R a_\mu \, \der_\rho \Xi^\mu_{\lam_{s+1} \dots \lam_{r+s}})
\\
&+ c^\R b_\rho \, \Xi^\rho_{\lam_1 \dots \lam_r} \,
\der_\R b c^\R a_\mu \, \Sig^\mu_{\lam_{r+1} \dots \lam_{r+s}}
- (- 1)^{rs} \,
c^\R b_\rho \, \Sig^\rho_{\lam_1 \dots \lam_s} \,
\der_\R b c^\R a_\mu \, \Xi^\mu_{\lam_{s+1} \dots \lam_{r+s}} +
\\
&- r \, c^\R a_\mu \, \Xi^\mu_{\lam_1 \dots \lam_{r-1} \rho} \,
\der_{\lam_r} \Sig^\rho_{\lam_{r+1} \dots \lam_{r+s}}
+ (- 1)^{rs} \, s \,
c^\R a_\mu \, \Sig^\mu_{\lam_1 \dots \lam_{s-1} \rho} \,
\der_{\lam_s} \Xi^\rho_{\lam_{s+1} \dots \lam_{r+s}}) \,
d^{\lam_1} \wed \dots \wed d^{\lam_{r+s}} \ten \der_\R a
\\[4mm]
&=
(\Xi^\rho_{\lam_1 \dots \lam_r} \,
\der_\rho \Sig^\mu_{\lam_{r+1} \dots \lam_{r+s}}
- (- 1)^{rs} \,
\Sig^\rho_{\lam_1 \dots \lam_s} \,
\der_\rho \Xi^\mu_{\lam_{s+1} \dots \lam_{r+s}}
\\
&-
r \, \Xi^\mu_{\lam_1 \dots \lam_{r-1} \rho} \,
\der_{\lam_r} \Sig^\rho_{\lam_{r+1} \dots \lam_{r+s}}
+ (- 1)^{rs} \, s \,
\Sig^\mu_{\lam_1 \dots \lam_{s-1} \rho} \,
\der_{\lam_s} \Xi^\rho_{\lam_{s+1} \dots \lam_{r+s}}) \,
d^{\lam_1} \wed \dots \wed d^{\lam_{r+s}} \ten \der_\mu
\\
&+
c^\R a_\mu \,
(\Xi^\rho_{\lam_1 \dots \lam_r} \,
\der_\rho \Sig^\mu_{\lam_{r+1} \dots \lam_{r+s}}
- (- 1)^{rs} \,
\Sig^\rho_{\lam_1 \dots \lam_s} \,
\der_\rho \Xi^\mu_{\lam_{s+1} \dots \lam_{r+s}}
\\
&-
r \, \Xi^\mu_{\lam_1 \dots \lam_{r-1} \rho} \,
\der_{\lam_r} \Sig^\rho_{\lam_{r+1} \dots \lam_{r+s}}
+ (- 1)^{rs} \, s \,
\Sig^\mu_{\lam_1 \dots \lam_{s-1} \rho} \,
\der_{\lam_s} \Xi^\rho_{\lam_{s+1} \dots \lam_{r+s}}) \,
d^{\lam_1} \wed \dots \wed d^{\lam_{r+s}} \ten \der_\R a
\\
&+
(\der_\rho c^\R a_\mu + c^\R b_\rho \, \der_\R b c^\R a_\mu) \,
(\Xi^\rho_{\lam_1 \dots \lam_r} \,
\Sig^\mu_{\lam_{r+1} \dots \lam_{r+s}} - (-1)^{rs}
\Sig^\rho_{\lam_1 \dots \lam_s} \,
\Xi^\mu_{\lam_{r+1} \dots \lam_{r+s}}) \,
d^{\lam_1} \wed \dots \wed d^{\lam_{r+s}} \ten \der_\R a
\\[4mm]
&=
(\Xi^\rho_{\lam_1 \dots \lam_r} \,
\der_\rho \Sig^\mu_{\lam_{r+1} \dots \lam_{r+s}}
- (- 1)^{rs} \,
\Sig^\rho_{\lam_1 \dots \lam_s} \,
\der_\rho \Xi^\mu_{\lam_{s+1} \dots \lam_{r+s}}
\\
&-
r \, \Xi^\mu_{\lam_1 \dots \lam_{r-1} \rho} \,
\der_{\lam_r} \Sig^\rho_{\lam_{r+1} \dots \lam_{r+s}}
+ (- 1)^{rs} \, s \,
\Sig^\mu_{\lam_1 \dots \lam_{s-1} \rho} \,
\der_{\lam_s} \Xi^\rho_{\lam_{s+1} \dots \lam_{r+s}}) \,
d^{\lam_1} \wed \dots \wed d^{\lam_{r+s}} \ten \der_\mu
\\
&+
c^\R a_\mu \,
(\Xi^\rho_{\lam_1 \dots \lam_r} \,
\der_\rho \Sig^\mu_{\lam_{r+1} \dots \lam_{r+s}}
- (- 1)^{rs} \,
\Sig^\rho_{\lam_1 \dots \lam_s} \,
\der_\rho \Xi^\mu_{\lam_{s+1} \dots \lam_{r+s}}
\\
&-
r \, \Xi^\mu_{\lam_1 \dots \lam_{r-1} \rho} \,
\der_{\lam_r} \Sig^\rho_{\lam_{r+1} \dots \lam_{r+s}}
+ (- 1)^{rs} \, s \,
\Sig^\mu_{\lam_1 \dots \lam_{s-1} \rho} \,
\der_{\lam_s} \Xi^\rho_{\lam_{s+1} \dots \lam_{r+s}}) \,
d^{\lam_1} \wed \dots \wed d^{\lam_{r+s}} \ten \der_\R a
\\
&+
(\der_\rho c^\R a_\mu + c^\R b_\rho \, \der_\R b c^\R a_\mu) \,
(\Xi^\rho_{\lam_1 \dots \lam_r} \,
\Sig^\mu_{\lam_{r+1} \dots \lam_{r+s}} -
\Xi^\mu_{\lam_1 \dots \lam_r}
\Sig^\rho_{\lam_{r+1} \dots \lam_{r+s}}) \,
d^{\lam_1} \wed \dots \wed d^{\lam_{r+s}} \ten \der_\R a \,.\QED
\end{align*}
\end{gather*}
\epf

For each
$\Psi \in \sec(\f E, \f Q) \,,$
we obtain the covariant differentials
\beq
\nab[c] \Psi \in (\f E, \, T^*\f E \ten \f Q)
\ssep{and}
d[c] \ti\Psi \in \sec(\f Q, \, T^*\f E \ten V\f Q) \,,
\eeq
with coordinate expressions
\beq
\nab[c] \Psi =
(\der_\lam \psi^\R a - c^\R a_\lam \com \, \Psi) \,
d^\lam \ten \E b_\R a
\ssep{and}
d[c] \ti\Psi =
(\der_\lam \psi^\R a - \der_\R bc^\R a_\lam \, \psi^\R b) \,
d^\lam \ten \der_\R a \,.
\eeq

Now, let us consider a (real) linear connection
$c \,.$

The above covariant differentials
$\nab[c] \Psi$
and
$d[c] \ti\Psi$
can be naturally identified.

The connection
$c$
turns out to be {\em complex linear\/} if and only if
$\nab (\coi \Psi) = \coi \nab \Psi \,,$
for each
$\Psi \in \sec(\f E, \f Q) \,.$

\bLm
$L(c) \E h : \f Q \to \Cn \ten (T^*\f E \ten V^*\f Q)$
and
$\nab \E h : \f E \to \Cn \ten (T^*\f E \ten \f Q^* \ten \f Q^*)$
turn out to be equal, up to a natural isomorphism.
\eLm

\bpf
We have the coordinate expressions
\bal
L(c) \, \E h
&=
\big(2 \, c^1_{\lam \, 1} \, w^1 +
(c^2_{\lam \, 1} + c^1_{\lam \, 2}) \, w^2
- \coi c^\R a_{\lam \, \R a} \, w^2
\big) \, d^\lam \ten \ch d^1
\\
&+
\big(2 \, c^2_{\lam \, 2} \, w^2 +
(c^2_{\lam \, 1} + c^1_{\lam \, 2}) \, w^1
+ \coi c^\R a_{\lam \, \R a} \, w^1
\big) \, d^\lam \ten \ch d^2
\\[3mm]
\nab(c) \, \E h
&=
d^\lam \ten \big(2 \, c^1_{\lam \, 1} \, w^1 +
(c^2_{\lam \, 1} + c^1_{\lam \, 2}) \, w^2
- \coi c^\R a_{\lam \, \R a} \, w^2
\big) \ten w^1
\\
&+
d^\lam \ten \big(2 \, c^2_{\lam \, 2} \, w^2 +
(c^2_{\lam \, 1} + c^1_{\lam \, 2}) \, w^1
+ \coi c^\R a_{\lam \, \R a} \, w^1
\big) \ten w^2 \,.\QED
\end{align*}
\epf

\bPr
The connection
$c$
turns out to be {\em Hermitian} (see also
\cite{GreHalVan72,Wel73}) if and only if
$\nab \, \E h = 0 \,,$
i.e. if and only if
\beq
d \big(\E h(\Psi, \, \Phi)\big) =
\E h\big(\nab\Psi, \, \Phi\big) +
\E h\big(\Psi, \, \nab\Phi\big) \,,
\qquad
\Al \, \Psi, \Phi \in \sec(\f E, \f Q) \,.
\eeq

According to
Proposition
\ref{expression of Hermitian tangent valued forms},
$c$
is Hermitian if and only if it is locally of the type
\beq
c = \chi[\E b] + \coi A[\E b] \ten \B I \,,
\ssep{with}
A[\E b] \in \sec(\f E, \, T^*\f E) \,.
\eeq

In other words,
$c$
is Hermitian if and only if
$c^1_{\lam 1} = c^2_{\lam 2} = 0$
and
$c^2_{\lam 1} = - c^1_{\lam 2} \,,$
i.e. if and only if its coordinate expression is of
the type
\beq
c = d^\lam \ten (\der_\lam + \coi A_\lam \, \B I) \,,
\ssep{with}
A_\lam = c^2_{\lam 1} \in \map(\f E, \Rn) \,.\ENDE
\eeq
\ePr

\smallskip

Now, let
$c$
be Hermitian.

We have the coordinate expression
$\nab \Psi =
(\der_\lam \psi - \coi A_\lam \, \psi) \, d^\lam \ten \E b \,,
\quad
\Al \, \Psi \in \sec(\f E, \f Q) \,.$

\bLm
For each
$\Xi \in \her(\f Q, \, \Lam^rT^*\f E \ten T\f Q) \,,$
we obtain
\beq
d[c]\Xi = \coi \br{\big(d[c] \, \Xi\big)\;} \ten \B I \,,
\eeq
where
$\br{\big(d[c] \, \Xi\big)\;} \in \sec(\f E, \, \Lam^{r+1}T^*\f E)$
is given by
\bal
\br{\big(d[c] \, \Xi\big)\;}
&=
L(\f1_\f E) \br\Xi - (-1)^r L(\ul\Xi) \, A[\E b]
\\
&=
d\br\Xi - (-1)^r L(\ul\Xi) \, A[\E b]
\end{align*}
and has coordinate expression
\beq
\br{\big(d[c] \, \Xi\big)\;} =
\big(\der_{\lam_1} \br \Xi_{\lam_2 \dots \lam_{s+1}}
- (A_\rho \, \der_{\lam_1} \Xi^\rho_{\lam_2 \dots \lam_{s+1}}
+ \der_\rho A_{\lam_1} \,
\Xi^\rho_{\lam_2 \dots \lam_{s+1}})
\big) \,
d^{\lam_1} \wed \dots \wed d^{\lam_{s+1}} \,.\ENDE
\eeq
\eLm

The curvature of
$c$
is
\beq
R[c] = - \coi \Phi[c] \ten \B I \,,
\eeq
where
$\Phi[c] : \f E \to \Lam^2T^*\f E$
is the closed 2--form given locally by
$\Phi[c] = 2 d A[\E b] \,.$

Thus, we have the coordinate expression
$\Phi[c] = 2 \, \der_\mu A_\lam \, d^\mu \wed d^\lam \,.$
%--------------------------------------------------------------------%
\myssec{Global classification}
\label{Global classification}
%--------------------------------------------------------------------%
\bsm
Eventually, we show that the choice of a Hermitian connection yields
a global classification of the Lie algebra of Hermitian tangent valued
forms of the line bundle.
\esm

Let us consider a Hermitian connection
$c \,.$

\smallskip

\bLm\label{horizontal hermitian vector fields}
If
$\ul\Xi \in \sec(\f E, \Lam^rT^*\f E \ten T\f E) \,,$
then
$c(\ul\Xi) \in \her(\f Q, \, \Lam^rT^*\f E \ten T\f Q) \,.$\ENDE
\eLm

\bPr\label{classification of hermitian tangent valued forms}
We have the following mutually inverse isomorphisms
\bal
\F h[c]
&:
\her(\f Q, \, \Lam^rT^*\f E \ten T\f Q) \to
\sec(\f E, \, \Lam^rT^*\f E \ten T\f E) \car
\sec(\f E, \, \Lam^rT^*\f E)
\\
\F j[c]
&:
\sec(\f E, \, \Lam^rT^*\f E \ten T\f E) \car
\map(\f E, \, \Lam^rT^*\f E) \to
\her(\f Q, \, \Lam^rT^*\f E \ten T\f Q) \,,
\end{align*}
given by
\beq
\F h[c] :
\Xi \mto \Big(\ul\Xi, \, - \coi \tr \big(\nu[c](\Xi)\big)\Big)
\ssep{and}
\F j[c] :
(\ul\Xi, \br \Xi) \mto c(\ul\Xi) + \coi \br \Xi \ten \B I \,,
\eeq
i.e., in coordinates
\bal
\F h[c] (\Xi)
&=
\big(\Xi^\mu_{\lam_1 \dots \lam_r} \,
d^{\lam_1} \wed \dots \wed d^{\lam_r} \ten \der_\mu \,,
\quad
(\br\Xi_{\lam_1 \dots \lam_r} -
A_\rho \, \Xi^\rho_{\lam_1 \dots \lam_r}) \,
d^{\lam_1} \wed \dots \wed d^{\lam_r}
\big)
\\
\F j[c] (\ul\Xi, \br \Xi)
&=
d^{\lam_1} \wed \dots \wed d^{\lam_r} \ten \big(
\Xi^\mu_{\lam_1 \dots \lam_r} \, \der_\mu +
\coi (A_\rho \, \Xi^\rho_{\lam_1 \dots \lam_r} +
\br\Xi_{\lam_1 \dots \lam_r}) \ten \B I
\big) \,.\ENDE
\end{align*}
\ePr

\bLm
Let us consider a closed 2-form
$\Phi$
of
$\f E$
and define the bracket
\bml
\big(\sec(\f E, \, \Lam^rT^*\f E \ten T\f E) \, \car \,
\sec(\f E, \, \Lam^rT^*\f E)\big)
\car
\big(\sec(\f E, \, \Lam^sT^*\f E \ten T\f E) \, \car \,
\sec(\f E, \, \Lam^sT^*\f E)\big)
\\
\to
\big(\sec(\f E, \, \Lam^rT^*\f E \ten T\f E) \, \car \,
\sec(\f E, \, \Lam^{r+s}T^*\f E)\big) \,,
\end{multline*}
given by
\beq
\big[(\ul\Xi_1, \br \Xi_1) \,, \; (\ul\Xi_2, \br \Xi_2)\big]_\Phi
\byd
\big([\ul\Xi_1, \ul\Xi_2] \,, \quad
\Phi (\ul\Xi_1, \ul\Xi_2) +
L(\ul\Xi_1) \, \br \Xi_2 - (-1)^{rs} L(\ul\Xi_2) \, \br \Xi_1
\big) \,,
\eeq
where
$\Phi (\ul\Xi_1, \ul\Xi_2)$
is defined, via decomposable tangent valued forms, as
\beq
\Phi(\xi \ten X, \sig \ten Y) \byd (\xi \wed \sig) \, \Phi(X, Y) \,.
\eeq

Then, the above bracket turns out to be a graded Lie bracket.
\eLm

\bpf
The graded commutativity of the 1st component follows from the fact
that
$[\ul\Xi_1, \ul\Xi_2]$
is the FN bracket, which is a graded Lie bracket.

Moreover, the anticommutativity of the 2nd component follows from the
equality
\beq
\Phi (\ul\Xi_1, \ul\Xi_2) +
L(\ul\Xi_1) \, \br \Xi_2 - (-1)^{rs} L(\ul\Xi_2) \, \br \Xi_1 =
- (-1)^{rs}
\big(\Phi (\ul\Xi_2, \ul\Xi_1) +
L(\ul\Xi_2) \, \br \Xi_1 - (-1)^{rs} L(\ul\Xi_1) \, \br \Xi_2) \,.
\eeq

\smallskip

Next, let us prove the Jacobi property.
Let us consider three pairs
$\Pi_i \byd (\ul\Xi_i, \br\Xi_i) \,,$
with
\beq
\ul\Xi_i \in \sec(\f E, \Lam^{\ba i}T^*\f E \ten T\f E)
\ssep{and}
\br\Xi_i \in \sec(\f E, \Lam^{\ba i}T^*\f E) \,,
\eeq
where
$\ba i$
denotes the degree of the $i$--th form, and set
\beq
(\ul\Sig, \br\Sig) \byd
\big[\Pi_1, \; [\Pi_2, \, \Pi_3]_\Phi\big]_\Phi
+ (-1)^{\ba 1 (\ba 2 + \ba 3)} \,
\big[\Pi_2, \; [\Pi_3, \, \Pi_1]_\Phi\big]_\Phi
+ (-1)^{\ba 3 (\ba 1 + \ba 2)} \,
\big[\Pi_3, \; [\Pi_1, \, \Pi_2]_\Phi\big]_\Phi \,,
\eeq
where
\beq
[\Pi_i,  \Pi_j]_\Phi
\byd
\big([\ul\Xi_i, \ul\Xi_j] \,,
\quad
\Phi (\ul\Xi_i, \ul\Xi_j) +
L(\ul\Xi_i) \, \br \Xi_j - (-1)^{\ba i \ba j} L(\ul\Xi_j) \, \br \Xi_i
\big) \,.
\eeq

Then, the Jacobi property of the 1st component follows from the
Jacobi property of the FN bracket
\beq
\ul\Sig \byd
\big[\ul\Xi_1, \; [\ul\Xi_2, \, \ul\Xi_3]\big]
+ (-1)^{\ba 1 (\ba 2 + \ba 3)} \,
\big[\ul\Xi_2, \; [\ul\Xi_3, \, \ul\Xi_1]\big]
+ (-1)^{\ba 3 (\ba 1 + \ba 2)} \,
\big[\ul\Xi_3, \; [\ul\Xi_1, \, \ul\Xi_2]\big] = 0 \,.
\eeq

Moreover, the Jacobi property of the 2nd component follows from the
following facts.

We have
\bal
\br\Sig
&=
\Phi\big(\ul\Xi_1, \, [\ul\Xi_2, \ul\Xi_3]\big)
+ (-1)^{\ba 1(\ba 3 + \ba 2)} \,
\Phi\big(\ul\Xi_2, \, [\ul\Xi_3, \ul\Xi_1]\big)
+ (-1)^{\ba 3(\ba 1 + \ba 2)} \,
\Phi\big(\ul\Xi_3, \, [\ul\Xi_1, \ul\Xi_2]\big)
\\
&+
L(\ul\Xi_1)\Phi(\ul\Xi_2, \ul\Xi_3) +
(-1)^{\ba 1(\ba 3 + \ba 2)} \,
L(\ul\Xi_2)\Phi(\ul\Xi_3, \ul\Xi_1) +
(-1)^{\ba 3(\ba 1 + \ba 2)} \,
L(\ul\Xi_3)\Phi(\ul\Xi_1, \ul\Xi_2)
\\
&+
\bigg(L(\ul\Xi_1)L(\ul\Xi_2) -
(-1)^{\ba 1 \ba 2} \,
L(\ul\Xi_2)L(\ul\Xi_1) -
L(\big[\ul\Xi_1\,,\; \ul\Xi_2\big]\bigg)\br\Xi_3
\\
&+
(-1)^{\ba 1 ( \ba 2 + \ba 3)} \,
\bigg(L(\ul\Xi_2)L(\ul\Xi_3) -
(-1)^{\ba 2 \ba 3} \,
L(\ul\Xi_3)L(\ul\Xi_2) -
L(\big[\ul\Xi_2\,,\; \ul\Xi_3\big]\bigg)\br\Xi_1
\\
&+
(-1)^{\ba 3(\ba 1 + \ba 2)}
\bigg(L(\ul\Xi_3)L(\ul\Xi_1) -
(-1)^{\ba 3 \ba 1} \,
L(\ul\Xi_1)L(\ul\Xi_3) -
L(\big[\ul\Xi_3\,,\; \ul\Xi_1\big] \bigg)\br\Xi_2
\\[3mm]
&=
\Phi(\ul\Xi_1, \, [\ul\Xi_2, \ul\Xi_3]) +
(-1)^{\ba 1(\ba 3 + \ba 2)} \,
\Phi(\ul\Xi_2, \, [\ul\Xi_3, \ul\Xi_1]) +
(-1)^{\ba 3(\ba 1 + \ba 2)} \,
\Phi(\ul\Xi_3, \, [\ul\Xi_1, \ul\Xi_2])
\\
&+
L(\ul\Xi_1)\Phi(\ul\Xi_2, \ul\Xi_3 ) +
(-1)^{\ba 1(\ba 3 + \ba 2)} \,
L(\ul\Xi_2)\Phi(\ul\Xi_3, \ul\Xi_1 ) +
(-1)^{\ba 3(\ba 1 + \ba 2)} \,
L(\ul\Xi_3)\Phi(\ul\Xi_1, \ul\Xi_2 ) \,.
\end{align*}

On the other hand, for decomposable tangent valued forms
$\ul\Xi_i=\xi_i \ten X_i$
we obtain
\bal
\br\Sig
&=
\Phi\big(X_1, \, \big[X_2, X_3\big]\big) \,
\xi_1 \wed \xi_2 \wed \xi_3 +
(-1)^{\ba 1(\ba 3+\ba 2)} \,
\Phi\big(X_2, \, \big[X_3, X_1\big]\big) \,
\xi_2 \wed \xi_3 \wed \xi_1
\\
&+
(-1)^{\ba 3(\ba 1+\ba 2)} \,
\Phi\big(X_3, \, \big[X_1, X_2\big]\big) \,
\xi_3 \wed \xi_1 \wed \xi_2
\\
&+
\Phi(X_1,X_3) \, \xi_1 \wed \xi_2 \wed L(X_2) \xi_3 +
(-1)^{\ba 1(\ba 3+\ba 2)} \,
\Phi(X_2,X_1) \, \xi_2 \wed \xi_3 \wed L(X_3) \xi_1
\\
&+
(-1)^{\ba 3(\ba 1+\ba 2)} \,
\Phi(X_3,X_2) \, \xi_3 \wed \xi_1 \wed L(X_1) \xi_2
\\
&-
(-1)^{\ba 2\ba 3} \,
\Phi(X_1,X_2) \, \xi_1 \wed\xi_3 \wed L(X_3) \xi_2 -
(-1)^{\ba 1\ba 2} \,
\Phi(X_2,X_3) \, \xi_2 \wed \xi_1 \wed L(X_1) \xi_3
\\
&-
(-1)^{\ba 1\ba 2+\ba 3(\ba 1+\ba 2)} \,
\Phi( X_3,X_1) \, \xi_3 \wed \xi_2 \wed L(X_2) \xi_1
\\
&+
(-1)^{\ba 2} \,
\Phi(X_1,X_3) \, \xi_1 \wed d\xi_2 \wed i(X_2) \xi_3 +
(-1)^{\ba 3 + \ba 1(\ba 3+\ba 2)} \,
\Phi(X_2,X_1) \, \xi_2 \wed d\xi_3 \wed i(X_3) \xi_1
\\
&+
(-1)^{\ba 1+\ba 3(\ba 1+\ba 2)} \,
\Phi(X_3,X_2) \,\xi_3 \wed d\xi_1 \wed i(X_1) \xi_2
\\
&-
(-1)^{\ba 3+\ba 2\ba 3} \,
\Phi( X_1,X_2) \, \xi_1 \wed d\xi_3 \wed i(X_3)\xi_2 -
(-1)^{\ba 1 + \ba 1\ba 2} \,
\Phi( X_2,X_3) \, \xi_2 \wed d\xi_1 \wed i(X_1)\xi_3
\\
&-
(-1)^{\ba 2+\ba 1\ba 2+\ba 3(\ba 1+\ba 2)} \,
\Phi( X_3,X_1) \, \xi_3 \wed d\xi_2 \wed i(X_2)\xi_1
\\
&+
\xi_1\wed L(X_1)\big(\Phi(X_2,X_3) \, \xi_2\wed \xi_3 \big)
+ (-1)^{\ba 1(\ba 3+\ba 2)} \,
\xi_2 \wed L(X_2)
\big(\Phi(X_3,X_1) \, \xi_3 \wed \xi_1 \big)
\\
&+
(-1)^{\ba 3(\ba 1+\ba 2)} \,
\xi_3\wed L(X_3)\big(\Phi(X_1,X_2) \, \xi_1\wed \xi_2\big)
\\
&+
(-1)^\ba 1
d\xi_1\wed i(X_1)\big(\Phi(X_2,X_3) \, \xi_2\wed \xi_3 \big)
+ (-1)^{\ba 2+\ba 1(\ba 3+\ba 2)} \,
d\xi_2\wed i(X_2)\big(\Phi(X_3,X_1) \, \xi_3\wed \xi_1 \big)
\\
&+
(-1)^{\ba 3+\ba 3(\ba 1+\ba 2)} \,
d\xi_3\wed i(X_3)\big(\Phi(X_1,X_2) \, \xi_1\wed \xi_2\big)
\,,
\end{align*}
i.e.
\bal
\br\Sig
&=
\bigg(
\Phi\big(X_1,\big[X_2\,\;X_3\big]\big) +
\Phi\big(X_2,\big[X_3\,\;X_1\big]\big) +
\Phi\big(X_3,\big[X_1\,\;X_2\big]\big)
\\
&\qquad+
X_1.\Phi(X_2,X_3) +
X_2.\Phi(X_3,X_1) +
X_3.\Phi(X_1,X_2)
\bigg) \, \xi_1\wed \xi_2 \wed \xi_3
\\
&+
\Phi(X_1,X_3)
\big(\xi_1 \wed \xi_2 \wed L(X_2) \xi_3 +
(-1)^{\ba 1\ba 2+\ba 3\ba 1+\ba 3\ba 2} \,
\xi_3 \wed \xi_2 \wed L(X_2) \xi_1
\\
&\qquad\qquad\qquad\;-
(-1)^{\ba 1\ba 3+\ba 1\ba 2} \,
\xi_2 \wed  L(X_2) (\xi_3 \wed \xi_1)\big)
\\
&+
\Phi(X_1,X_2)
\big(-(-1)^{\ba 1\ba 3 +\ba 1\ba 2} \,
\xi_2 \wed \xi_3 \wed L(X_3) \xi_1 -
(-1)^{\ba 3\ba 2} \,
\xi_1 \wed \xi_3 \wed L(X_3) \xi_2
\\
&\qquad\qquad\qquad\;+
(-1)^{\ba 1\ba 3+\ba 3\ba 2} \,
\xi_3 \wed  L(X_3) (\xi_1 \wed \xi_2)\big)
\\
&+
\Phi(X_2,X_3)
\big(-(-1)^{\ba 1\ba 3 + \ba 3 \ba 2} \,
\xi_3 \wed \xi_1 \wed L(X_1) \xi_2 -
(-1)^{\ba 1\ba 2} \,
\xi_2 \wed \xi_1 \wed L(X_1) \xi_3
\\
&\qquad\qquad\qquad\; +
\xi_1 \wed  L(X_1) (\xi_2 \wed \xi_3)\big)
\\
&+
\Phi(X_1,X_3)
\big((-1)^\ba 2\, \xi_1 \wed d\xi_2 \wed i(X_2) \xi_3 +
(-1)^{\ba 2+\ba 1\ba 2+\ba 3\ba 1+\ba 3\ba 2} \,
\xi_3 \wed d\xi_2 \wed i(X_2) \xi_1
\\
&\qquad\qquad\qquad\;-
(-1)^{\ba 2+ \ba 1\ba 3+\ba 1\ba 2} \,
d\xi_2 \wed  i(X_2) (\xi_3 \wed \xi_1)\big)
\\
&+
\Phi(X_1,X_2)
\big(-(-1)^{\ba 3+ \ba 1\ba 3 +\ba 1\ba 2} \,
\xi_2 \wed d\xi_3 \wed i(X_3) \xi_1 -
(-1)^{\ba 3+\ba 3\ba 2} \,
\xi_1 \wed d\xi_3 \wed i(X_3) \xi_2
\\
&\qquad\qquad\qquad\;+
(-1)^{\ba 3+\ba 1\ba 3+\ba 3\ba 2} \,
d\xi_3 \wed  i(X_3) (\xi_1 \wed \xi_2)\big)
\\
&+
\Phi(X_2,X_3)
\big(-
(-1)^{\ba 1+\ba 1\ba 3 +\ba 3\ba 2} \,
\xi_3 \wed d\xi_1 \wed i(X_1) \xi_2 -
(-1)^{\ba 1+\ba 1\ba 2} \,
\xi_2 \wed d\xi_1 \wed i(X_1) \xi_3
\\
&\qquad\qquad\qquad\;+
(-1)^\ba 1
d\xi_1 \wed  i(X_1) (\xi_2 \wed \xi_3)\big)
\\[3mm]
&=
d\Phi(X_1,X_2,X_3) \, \xi_1 \wed \xi_2 \wed \xi_3 \,,
\end{align*}
which vanishes for a closed
$\Phi \,.$\QED
\epf

\smallskip

Now, let us refer to the 2--form
$\Phi[c] \byd \coi \tr R[c]$
associated with the curvature of
$c \,.$

\bTh\label{Lie algebra classification of hermitian vector fields}
The map
\beq
\F j[c] :
\sec(\f E, \, \Lam^rT^*\f E \ten T\f E) \car
\map(\f E, \, \Lam^rT^*\f E) \to
\her(\f Q, \, \Lam^rT^*\f E \ten T\f Q)
\eeq
is a graded Lie algebra isomorphism with respect to the graded Lie
bracket
$[\,,]_{\Phi[c]}$
and the FN bracket.
\eTh

\bpf
We have
\bal
[c(\ul\Xi), \, c(\ul\Sig)]
&= c\big([\ul\Xi, \, \ul\Sig]\big) - R[c](\ul\Xi, \ul\Sig) =
c\big([\ul\Xi, \, \ul\Sig]\big) +
\coi \Phi[c](\ul\Xi, \, \ul\Sig) \, \B I \,,
\\
\big[c(\ul\Xi), \, \coi \br\Sig \, \B I\big]
&= \coi (L(\ul\Xi) \, \br\Sig) \, \B I \,,
\\
\big[c(\ul\Sig), \, \coi \br \Xi \, \B I\big]
&= \coi (L(\ul\Sig) \, \br\Xi)\, \B I \,,
\\
[\coi \br\Xi \, \B I, \, \coi \br\Sig \, \B I]
&= 0 \,,
\end{align*}
which implies
\bal
\big[\F j[c] (\ul\Xi, \br\Xi) \,, \; \F j[c] (\ul\Sig, \br\Sig]\big]
&=
\big            [c(\ul\Xi) + \coi \br\Xi \, \B I \,, \quad
c(\ul\Sig) + \coi \br\Sig \, \B I\big]
\\
&=
\big            [c(\ul\Xi), \; c(\ul\Sig)] +
\big[c(\ul\Xi), \; \coi \br\Sig \, \B I\big] +
\big[\coi \br\Xi \, \B I, \; c(\ul\Sig)\big] +
\big            [\coi \br\Xi \, \B I, \; \coi \br\Sig \, \B I\big]
\\
&= c([\ul\Xi, \, \ul\Sig]) +
\coi \big(\Phi[c](\ul\Xi, \, \ul\Sig) +
L(\ul\Xi) \, \br\Sig - (-1)^{rs} \, L(\ul\Sig) \, \br\Xi\big)
\, \B I
\\
&=
\F j[c] \big([\ul\Xi, \, \ul\Sig] \,, \;\;
\Phi[c](\ul\Xi, \, \ul\Sig) +
L(\ul\Xi) \, \br\Sig - (-1)^{rs} \, L(\ul\Sig) \, \br\Xi\big)
\\
&= \F j[c] \big([(\ul\Xi, \br\Xi) \,, \;
(\ul\Sig, \br\Sig)]_{\Phi[c]}\big)
\,.\QED
\end{align*}
\epf

\bCr
The map
\beq
\her(\f Q, \, \Lam^rT^*\f E \ten T\f Q) \to
\sec(\f E, \, \Lam^rT^*\f E \ten T\f E) : \Xi \mto \ul\Xi
\eeq
is a central extension of  graded Lie algebras by
$\sec(\f E, \, \Lam^rT^*\f E) \,.$\ENDE
\eCr

\smallskip

We stress that in the Galilei and Einstein frameworks the graded Lie
bracket
$[\,,]_\Phi$
is closely related to a special phase bracket
\cite{JanMod99,JanMod97b}
induced by the background geometric structure of the base space
$\f E \,.$
Indeed, this is the source of a quantisation of forms.
But this matter is beyond the scope of the present purely geometric
paper.
%--------------------------------------------------------------------%
\bfz
%--------------------------------------------------------------------%

\efz
%--------------------------------------------------------------------%
\end{document}